# New interaction potentials for alkali and alkaline-earth aluminosilicate glasses

Siddharth Sundararaman[a,b], Liping Huang[a,†], Simona Ispas[b] and Walter Kob[b]

[a]Department of Materials Science and Engineering, Rensselaer Polytechnic Institute, Troy, NY 12180, United States

[b]Laboratoire Charles Coulomb (L2C), University of Montpellier, CNRS, F34095 Montpellier, France

**Abstract**

We apply a recently developed optimization scheme to obtain effective potentials for alkali and alkaline-earth aluminosilicate glasses that contains lithium, sodium, potassium, or calcium as modifiers. As input data for the optimization, we used the radial distribution functions of the liquid at high temperature generated by means of *ab initio* molecular dynamics simulations and density and elastic modulus of glass at room temperature from experiments. The new interaction potentials are able to reproduce reliably the structure and various mechanical and vibrational properties over a wide range of compositions for binary silicates. We have tested these potentials for various ternary systems and find that they are transferable and can be mixed, thus allowing to reproduce and predict the structure and properties of multi-component glasses.

## 1. Introduction

Alkali and alkaline-earth aluminosilicate glasses are of both great technological and scientific importance in various fields, but for most of these materials the connection between the atomic structure and macroscopic properties is not well understood, which makes it difficult to improve their properties[1–5]. Molecular dynamics (MD) is an excellent tool to predict structure and a variety of properties of such glasses and thus provides great insight into the structure-property relations. However, the accuracy and reliability of these predictions by MD simulations depend strongly on the interaction potential used[6–8]. The key element of such studies is hence the availability of interaction potentials that are reliable and transferrable over large compositional ranges and under different thermodynamic conditions[6].

Great efforts have been devoted over decades for developing potentials for both pure silica [9–19] and silicate-based glasses[12,20–26]. Early work to study the structure of alkali silicates in MD was done by Soules et al. on sodium silicate[27], by using a Born-Mayer functional form for short-range interactions. This functional form was subsequently used to study glass systems like borates, aluminosilicates and borosilicates[28–31]. The potentials for silica glass developed by Van-Beest et al. (BKS)[9] or Vessal et al. (VSL)[19] were used in conjunction with those developed for binary oxides[32,33] to study the structure and properties of various multi-component silicate glasses[34–37]. Other potentials developed to simulate properties of various multi-component oxide glasses include those parameterized by Teter[21–23] and by Delaye and coworkers[38–40]. These potentials used a Buckingham functional form for the short-range interactions, and the potentials by VSL and Delaye also include an angular three-body term[19,39]. All of these interaction potentials used fixed effective charges for the Coulombic interactions except the one by VSL, which used formal charges on the ions. Based on fixed effective charges, Pedone et al. developed a set of potentials for various glasses[12] using the Morse form for short-range interactions by fitting to structural and elastic properties of various oxide and silicate crystals. The availability of parameters for a large number of oxides and its ability to predict the mechanical properties fairly accurately[12,41] have made it a very popular choice for the study of multi-component oxide glasses[42–46].

---

†Corresponding author. E-mail address: huangL5@rpi.edu (L. Huang).

An interesting approach was proposed by Habasaki et al. who developed a set of potentials for alkali silicates by using the Buckingham function for short-range interactions and a constant effective charge for the cations but an oxygen charge which depends on the composition to partially emulate the polarization effect[26]. This set of potentials has the ability to predict well the trends in elastic moduli and density for some compositions although the absolute values were too low[47]. Polarizable potentials like the polarizable ion model (PIM)[48] and the aspherical ion model (AIM)[20] that account for anion polarization and shape deformations have also been used to study various silicate systems.

One of the most common problems with many of the existing potentials is their inability to quench glass of a reasonable density under normal conditions, i.e., in an NPT (constant number of atoms, constant pressure and constant temperature) ensemble, similar to the melt-quench process in experiments. For example, the potentials developed by Pedone et al. tend to highly overestimate the density of glass if quenched in the NPT ensemble. An NVT (constant number of atoms, constant volume and constant temperature) ensemble needs to be employed to produce a reasonable glass density at room temperature and even then it can sometimes produce large holes in the quenched samples[46]. Furthermore, potentials that include polarization effects or three-body interactions have the added disadvantage of being computationally very expensive.

In our previous work, we developed a new scheme to optimize interaction potential for silica glass that shows improved structure and properties under various thermodynamic conditions[6]. In the present work, we use a similar optimization scheme to develop pair potentials for various multi-component glasses that include alkali modifiers lithium, sodium and potassium, alkaline earth modifier calcium, and aluminum that, depending on composition can behave as a modifier or a former[49,50]. One of the major aims of this work is to produce computationally efficient force fields essential for large scale virtual mechanical tests of glass under complex loading conditions[51–53]. This is achieved by maintaining a simple pair functional form and using the composition dependent anion charge scheme suggested by Habasaki et al. to partially take into account the polarizability effect[26]. As we will demonstrate, these new interaction potentials can predict very well the melt structure as compared to *ab-initio* data and various structural and elastic properties of glasses over a wide range of compositions as compared to experimental data. We further show that these potentials are transferable and can be mixed to reproduce the structure and properties of various multi-component alkali and alkaline-earth aluminosilicate glasses.

The organization of the paper is as follows: In Sec. 2, we describe the details of how the optimization was adapted to include the various new elements. In Sec. 3, we will compare our new potentials with both *ab-initio* data and experimental data over a large range of compositions and demonstrate their reliability and transferability. Finally in Sec. 4 we will summarize our results and draw conclusions.

## 2. Simulation methods

In this section, we will briefly describe the optimization scheme developed for silica glass[6] and how we adapted this approach to include other elements.

### 2.1 Potential and cost function for the optimization

Following the work for silica glass[6], we used the Buckingham[54] functional form for the short-range interactions and evaluated the long-range Coulomb interactions by means of Wolf truncation method[55,56], i.e.,

$$V^{Buck}(r_{\alpha\beta}) = A_{\alpha\beta} exp(-B_{\alpha\beta} r_{\alpha\beta}) - \frac{C_{\alpha\beta}}{r_{\alpha\beta}^6} + \frac{D_{\alpha\beta}}{r_{\alpha\beta}^{24}} + V^W(r_{\alpha\beta}) \quad (1)$$

where

$$V^W(r_{\alpha\beta}) = q_\alpha q_\beta \left[ \frac{1}{r_{\alpha\beta}} - \frac{1}{r_{cut}^W} + \frac{(r_{\alpha\beta} - r_{cut}^W)}{(r_{cut}^W)^2} \right] \quad (2)$$

and $\alpha, \beta \in \{O, Si, Li, K, Na, Ca, Al\}$. The short-range interactions are truncated at 8 Å and $V^W$ at 10 Å. The repulsive term $\frac{D_{\alpha\beta}}{r_{\alpha\beta}^{24}}$ is added to avoid the divergence of the potential due to the van der Waals term at very small distances. In the following the short-range interaction parameters for O-O, O-Si and Si-Si pairs as well as silicon charge are kept constant during the optimization, i.e., are fixed to the values from our previous work[6], while the other interaction parameters are allowed to change. To impose charge neutrality, the oxygen charge is evaluated as follows[26]:

$$q_O = \frac{(1 - y_X - y_{Al})q_{Si} + 2y_X q_X + 2y_{Al} q_{Al}}{y_X - y_{Al} - 2}; \quad X \in Li, Na, K \quad (3)$$

$$q_O = \frac{(1 - y_{Ca} - y_{Al})q_{Si} + y_{Ca} q_{Ca} + 2y_{Al} q_{Al}}{y_{Ca} - y_{Al} - 2} \quad (4)$$

Eq. (3) is for alkali aluminosilicates and Eq. (4) is for alkaline-earth aluminosilicates, where $y_\alpha$ and $q_\alpha$ are the mole fraction and charge of the species $\alpha$, respectively. All classical MD simulations were carried out using the LAMMPS software[57] with a time step of 0.8 fs. The cost function for optimizing the parameters is given by

$$\chi^2(\phi) = w_1 \int_0^{r_{N_{RDF}}} \sum_{\alpha,\beta} (rg_{\alpha\beta}^{calc,3500K}(r|\phi) - rg_{\alpha\beta}^{ref,3500K}(r))^2 dr \\ + w_2 (E^{calc,300K}(\phi) - E^{ref,300K})^2 + w_3 (\rho^{calc,300K}(\phi) - \rho^{ref,300K})^2 \quad (5)$$

where $\phi$ is the current parameter set, $\alpha, \beta$ are the different species, $w_1, w_2, w_3$ are the weights for each contribution, $rg_{\alpha\beta}(r)$ is the radial distribution function (RDF) weighted by the distance $r$ up to a maximum distance of $r_{N_{RDF}} = 8\,\text{Å}$ at 3500 K, $\rho$ is the density and $E$ is the Young's modulus at 300 K and 0 GPa pressure. The superscript "ref" refers to the first principles or experimental reference data towards which the optimization was carried out, and superscript "calc" refers to the calculated properties using the current parameter set. Here the RDFs of the liquid were evaluated at high temperature while the density and Young's modulus were calculated at room temperature. We found that including the structure and properties at a single composition instead of multiple compositions was sufficient for obtaining a reliable dependence of the potentials on composition. It should be pointed out that pure silica is effectively an additional composition in the cost function during the optimization as we do not change those parameters. In some of the optimizations we had to use different compositions for the structure at high temperature and the properties at room temperature because of the unavailability of the experimental data, which adds to the compositional training. Finally we note that in the present study we did not include the vibrational density of states (VDOS) in the cost function as we did in our prior work for silica glass[6], and the reason for this will be given in Sec. 3.

The RDFs for "calc" were calculated by equilibrating a sample of about 1500 atoms at 3500 K at the glass density for 20 ps in the NVT ensemble, followed by a production run of 40 ps. For the various alkali systems, we used the composition $0.2X_2O$–$0.8SiO_2$ (X ∈ Li, Na, K) as the reference system at high temperature, while for the alkaline-earth system we considered $0.4CaO$–$0.6SiO_2$, and for the aluminosilicate melt we used $0.18Na_2O$–$0.18Al_2O_3$–$0.64SiO_2$, a composition that has been studied in previous MD simulations[58]. The samples at room temperature needed for the optimization were produced using a multistep process by first quenching samples of about 10000 atoms in the NVT ensemble at the glass density using the Pedone potential[12] followed by relaxation in the NPT ensemble at 300 K and zero pressure for 200 ps. These samples were used in an optimization step that included only the high temperature structure and density in the cost function to produce a first starting potential. The so-obtained potential was then used to quench liquids to 300 K at zero pressure in the NPT ensemble to generate glasses. These samples were used in later optimizations that also included the Young's modulus in the cost function. This process saves computational cost since there is no need of quenching samples to room temperature after each optimization step. This is justified as it was previously observed that the mechanical properties are more strongly dependent on the interaction potential than the structure of glass[59].

The density at room temperature was measured during the optimization by relaxing the quenched samples in the NPT ensemble at zero pressure and 300 K with the current parameter set. The Young's modulus $E_x$ was then measured by compressing and expanding the samples at 300 K along one direction at a constant strain rate (1.25/ns) up to a linear change of 0.6% and measuring their stress response:

$$E_x = \frac{d\sigma_x}{d\varepsilon_x} \tag{6}$$

Where $\sigma_x$ and $\varepsilon_x$ are the stress and strain, respectively, along the $x$ direction.

The reference compositions used for the glass properties for the alkali and alkaline-earth systems were $0.4X_2O$–$0.6SiO_2$ (X ∈ Li, Na, K), $0.5CaO$–$0.5SiO_2$ and $0.25Na_2O$–$0.25Al_2O_3$–$0.5SiO_2$, respectively. The minimization was performed using the Levenberg-Marquardt algorithm[60,61]. The final optimized effective charges and short-range interaction parameters for the different systems are given in Table 1 and Table 2, respectively. In the rest of the paper these new potentials will be referred to as "SHIK".

**Table 1:** Charge for different species.

| Species | Si | Na | K | Li | Ca | Al |
|---|---|---|---|---|---|---|
| Charge (e) | 1.7755 | 0.6018 | 0.6294 | 0.5727 | 1.3967 | 1.6334 |

**Table 2:** Short-range interaction parameters.

| i-j | $A_{ij}$ (eV) | $B_{ij}$ (Å$^{-1}$) | $C_{ij}$ (eV·Å$^6$) | $D_{ij}$ (eV·Å$^{24}$) |
|---|---|---|---|---|
| O-O | 1120.5 | 2.8927 | 26.132 | 16800 |
| O-Si | 23108 | 5.0979 | 139.70 | 66 |
| Si-Si | 2798.0 | 4.4073 | 0.0 | 3423204 |
| O-Na | 1127566 | 6.8986 | 40.562 | 16800 |
| Si-Na | 495653 | 5.4151 | 0.0 | 16800 |
| Na-Na | 1476.9 | 3.4075 | 0.0 | 16800 |
| O-K | 219750 | 5.2494 | 111.97 | 16800 |
| Si-K | 550659 | 4.8283 | 0.0 | 16800 |
| K-K | 1177.8 | 2.7363 | 0.0 | 16800 |
| O-Li | 6745.2 | 4.9120 | 41.221 | 70 |
| Si-Li | 17284 | 4.3848 | 0.0 | 16800 |
| Li-Li | 2323.8 | 3.9129 | 0.0 | 3240 |
| O-Ca | 101376 | 5.2914 | 68.684 | 16800 |
| Si-Ca | 544421 | 5.3256 | 0.0 | 16800 |
| Ca-Ca | 12384 | 5.4829 | 0.0 | 16800 |
| O-Al | 21740 | 5.3054 | 65.815 | 66 |
| Al-Al | 1799.1 | 3.6778 | 100.0 | 16800 |

Even though sodium aluminosilicate compositions were used as reference to optimize the parameters for the interactions involving aluminum, it is important to note that our tests showed that it is not necessary to include short-range Na-Al or Si-Al interactions. The reason for this is probably due to the fact that aluminum has an intermediate behavior and, depending on the composition[49,50], can act as both a network former and a modifier. This absence of Na-Al interactions allows us to simulate compositions with other alkali and alkaline earth elements without having to perform further optimizations, see Sec. 3 for details.

**2.2 Generation of reference data**

*Ab initio* MD simulations were carried out using the Vienna ab initio simulation package (VASP)[62,63]. The electronic structure was described by means of the Kohn–Sham formulation of the density functional theory[64] using the generalized gradient approximation (GGA) and the PBEsol functional[65,66]. The Kohn-Sham orbitals were expanded in a plane-wave basis at the Γ point of the supercell, containing components with energies up to 600 eV, and the electron-ion interaction was described using the projector-augmented-wave formalism[67,68]. For the solution of the Kohn-Sham equations, the residual minimization method-direct inversion in the iterative space[63] was chosen, and the electronic convergence criterion was fixed at $5\times10^{-7}$ eV. The choice of the above mentioned approximations and parameters have been motivated by previous *ab initio* studies of liquid and glassy states of pure silica[6] as well as sodium borosilicate compositions[69].

The *ab initio* MD simulations were carried out in the NVT ensemble at 3500 K using the Nosé thermostat[70] to control the temperature and by starting from configurations obtained from equilibrium classical MD simulations at the same temperature. A cubic system of N atoms with periodic boundary conditions was used with the simulation box length fixed to a value corresponding to a density close to the experimental value at ambient conditions for each composition[1,4] (see Table 3 for details). The simulation for a given composition was stopped once the mean squared displacement (MSD) of the slowest element, i.e., silicon, reached ~10 Å$^2$, which was sufficient for other species to reach the diffusive regime too. In each case, we discarded the first 1 to 2.5 ps of the trajectory in each case and used the remaining data for calculating the RDFs.

**Table 3:** Number of atoms and density used to equilibrate the liquid at high temperatures in *ab initio* MD simulations.

| System | N (atoms) | $\rho$ (g/cm$^3$) |
|---|---|---|
| 0.2Li$_2$O–0.8SiO$_2$ | 420 | 2.280 |
| 0.2Na$_2$O–0.8SiO$_2$ | 420 | 2.383 |
| 0.2K$_2$O–0.8SiO$_2$ | 420 | 2.389 |
| 0.4CaO–0.6SiO$_2$ | 390 | 2.779 |
| 0.18Na$_2$O–0.18Al$_2$O$_3$–0.64SiO$_2$ | 504 | 2.440 |

## 2.3 Glass preparation

To show the reliability and transferability of the new potentials, glasses of various compositions, shown in Table 4, were prepared using the melt-quench method. For this we equilibrated samples with 10500-12000 atoms at the experimental glass density for the composition at a temperature T$_1$ between 3500-4000 K for about 200 ps in the NVT ensemble (see Table 4). The samples were then cooled down to a second temperature T$_2$ between 2500-3000 K, equilibrated for 200 ps and subsequently quenched to 300 K in the NPT ensemble at a nominal quench rate of ~2.25 K/ps. A small pressure of ~100 MPa was applied at high temperature (T$_2$) which was ramped down to 0 GPa during the quenching process. The samples were then annealed at 300 K and 0 GPa for 100 ps in the NPT ensemble. Four independent samples were quenched for each composition to improve the statistics of the results.

**Table 4:** Details of the quenching process for each glass system.

| Composition | N (atoms) | T$_1$ (K) | T$_2$ (K) |
|---|---|---|---|
| yX$_2$O–(1-y)SiO$_2$ (X ∈ Na, K , y ∈ 0.1, 0.15, 0.2, 0.25, 0.3) | 10500 | 4000 | 3000 |
| yX$_2$–(1-y)SiO$_2$ (X ∈ Li , y ∈ 0.1, 0.15, 0.2, 0.25, 0.3) | 10500 | 3500 | 2500 |
| 0.1Na$_2$O–yCaO–(0.9-y)SiO$_2$ (y ∈ 0.1, 0.15, 0.2, 0.25, 0.3) | ~11500 | 3500 | 2500 |
| (0.25-y)Na$_2$O–yLi$_2$O–0.75SiO$_2$ | 12000 | 3500 | 2500 |
| (y ∈ 0.05, 0.1, 0.15, 0.2) | | | |
| 0.15X$_2$O–yAl$_2$O$_3$–(0.85-y)SiO$_2$ (X ∈ Na, Li, y ∈ 0.05, 0.1, 0.15, 0.2, 0.25) | ~10500 | 3500 | 3000 |

## 3. Results and Discussion

In this section, we will discuss the ability of the new potentials to reproduce *ab-initio* and experimental data in both the liquid and the glass state.

### 3.1 Structure of the liquid

We first compare the structure of the liquid as predicted by these new potentials with the one obtained from *ab initio* MD simulations. This comparison is useful since it allows to test whether the chosen functional form of the potential is indeed able to produce a reliable *equilibrium* structure, i.e., there are no issues with cooling rate dependence as it will be the case for a glass. Figure 1(a-c) show some of the partial RDFs predicted by these potentials at 3500 K for alkali/alkaline-earth silicates in comparison with data from *ab initio* simulations.

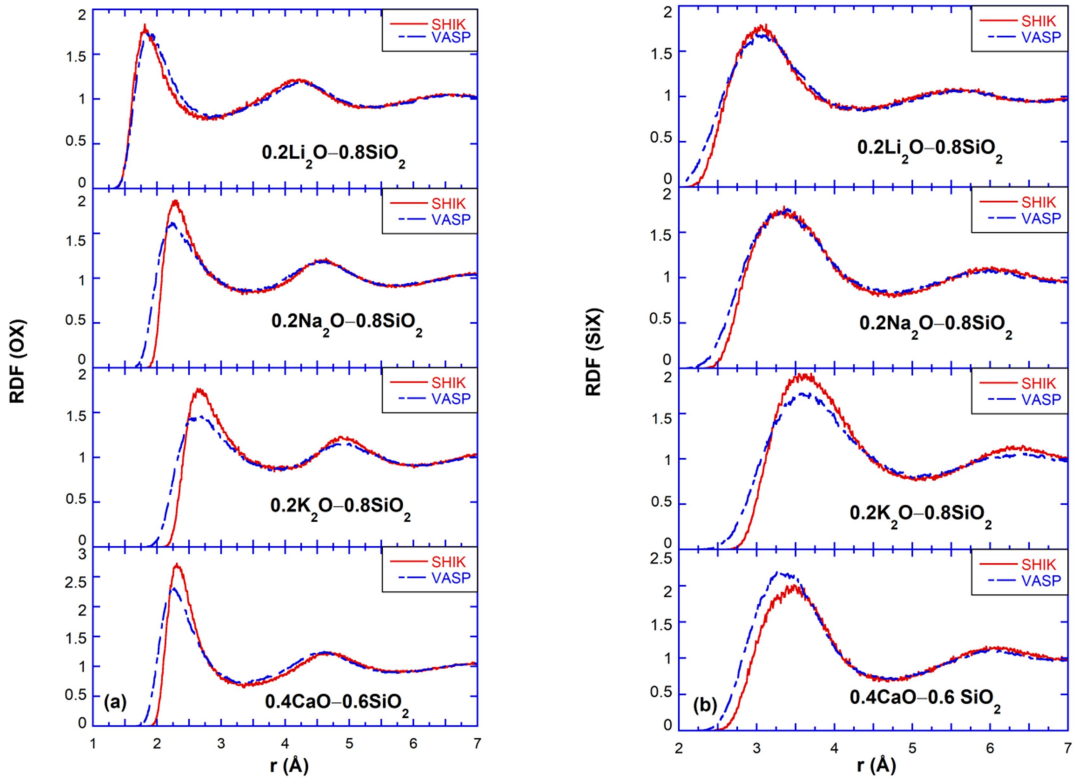

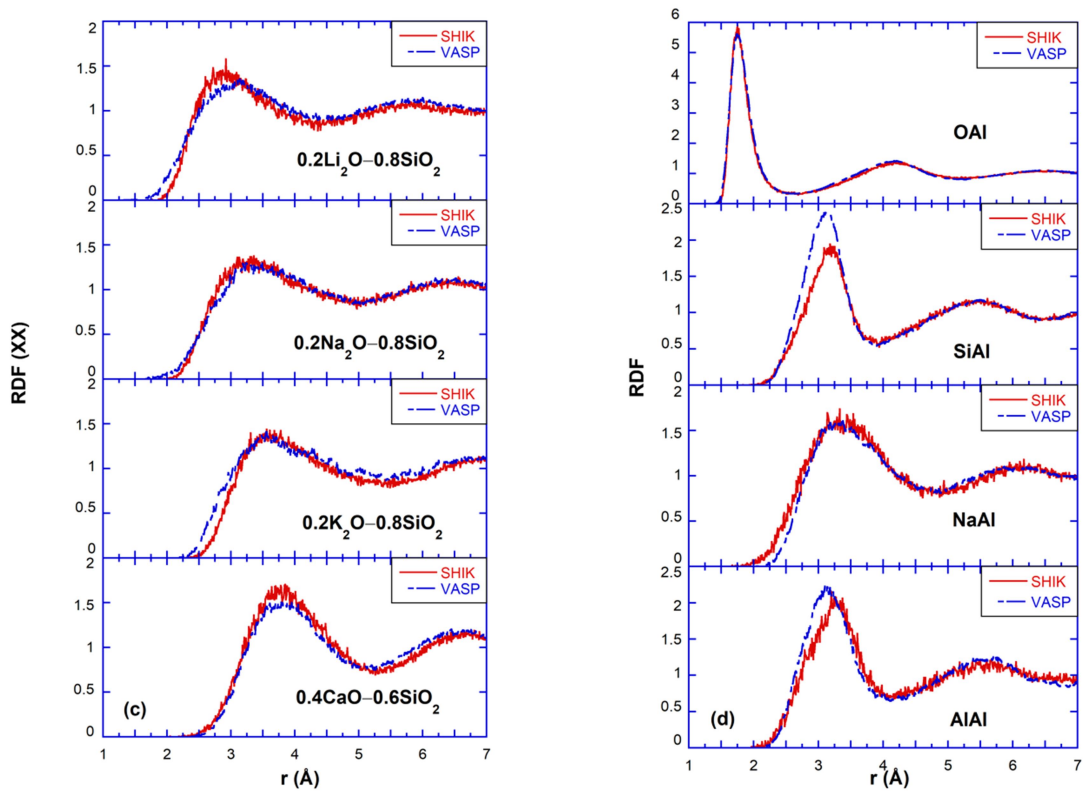

**Fig. 1** Partial radial distribution functions of the melt as obtained from the SHIK potential (red solid line) and *ab initio* simulations (blue dashed line) at 3500 K. (a) O-X, (b) Si-X and (c) X-X pair, where X is Li, Na, K and Ca, (d) O-Al, Si-Al, Na-Al and Al-Al pairs in $0.18Na_2O$–$0.18Al_2O_3$–$0.64SiO_2$.

Overall the new potentials are able to reproduce very well the structure of the melt of the various systems predicted by *ab-initio* simulations. This is not that surprising since these RDFs were included in the cost function that was minimized. Figure 1(a) shows that classical MD simulations are able to predict very well the position of the first nearest neighbor O-X peak as a function of composition but tend to predict more order than *ab initio* data in that the intensity of the peak is higher (X is Li, Na, K and Ca). Figure 1(d) shows various partial RDFs predicted for the ternary system $0.18Na_2O$–$0.18Al_2O_3$–$0.64SiO_2$ at 3500 K compared to *ab initio* data. Even though the O-Al and Na-Al partial RDFs are reproduced well, larger discrepancies are observed in the Al-Al and Si-Al partial RDFs. This is because we had to make a compromise between an accurate description of the latter two RDFs and other glass properties, as improving one of them during the optimization while constraining the glass density inadvertently deteriorated the other. We note that we observed no significant improvement even if the short-range Si-Al interactions were included during the optimization.

Figure 2(a-b) show the O-X-O and the Si-O-X bond angle distributions (BADs) for the various alkali and alkaline-earth compositions (X is Li, Na, K and Ca), while Fig. 2(c) shows the BADs

for the O-Al-O, Al-O-Al, Al-O-Si and Al-O-Na triplets in the aluminosilicate melt at 3500 K as compared to *ab initio* data.

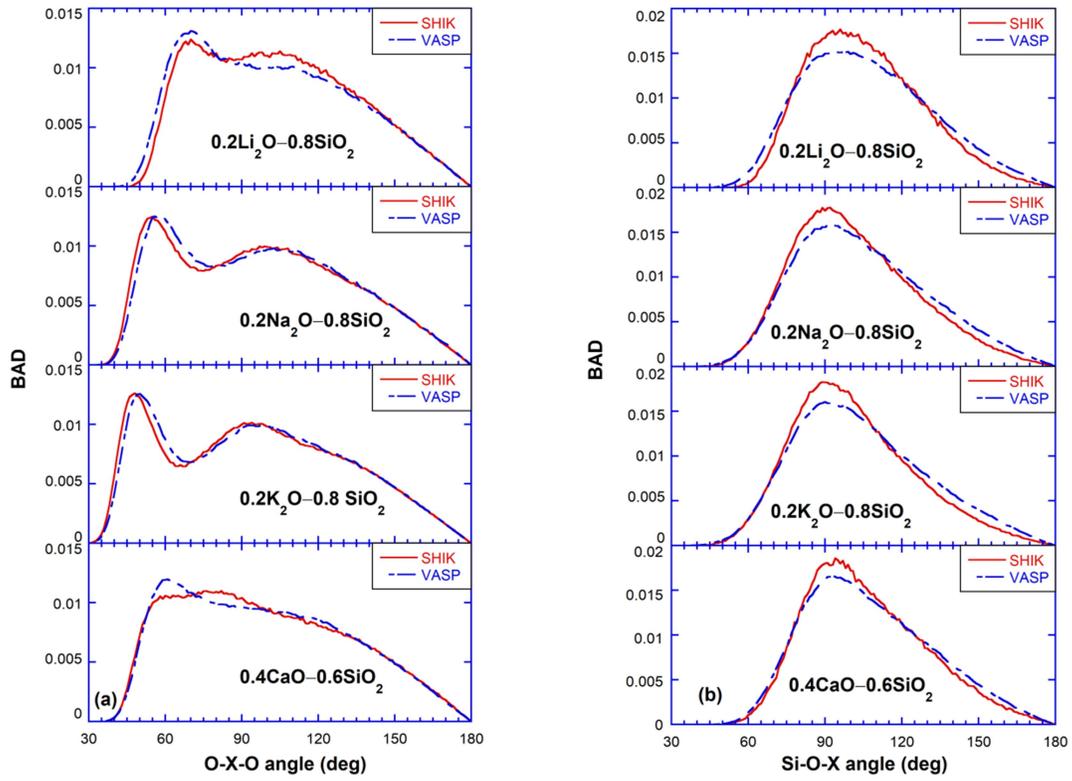

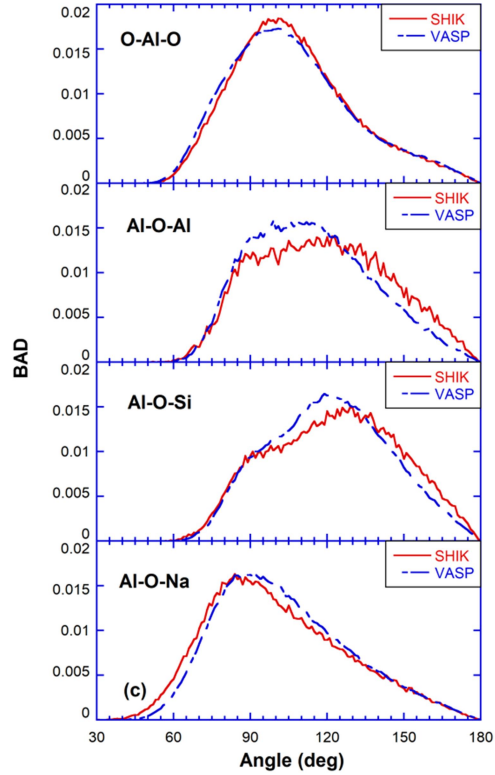

**Fig. 2** Bond angle distributions as obtained from the SHIK potential (red solid line) and *ab initio* simulations (blue dashed line) of the melt at 3500 K. (a) O-X-O, (b) Si-O-X, where X is Li, Na, K and Ca, (c) O-Al-O, Al-O-Al, Al-O-Si and Al-O-Na distributions in $0.18Na_2O$–$0.18Al_2O_3$–$0.64\ SiO_2$.

As we have already noted in our previous work[6], even though the BADs depend strongly on the RDFs that were included in the cost function, they are not entirely defined by the latter either and hence they can be considered as a quantity that gives new information. The difference observed in the O-X-O BAD predicted by classical and *ab initio* simulations can be attributed to the discrepancies observed in O-X and Si-X partial RDFs in Fig. 1. Figure 2(c) shows that similar to the case of silica glass, the intra-tetrahedral angle (O-Al-O) is predicted very well while the inter-tetrahedral angles (Al-O-Al and Al-O-Si) are overestimated as compared to *ab initio* data[6]. It is important to note that the discrepancies observed in the structure are not entirely due to the shortcomings of the pair potential functional form, but also due to compromises in the optimization to predict different properties reliably over a wide range of compositions.

### 3.2 Structure of the glass

The structure of the glass obtained from the SHIK potential can be compared to experiments by looking at the structure factor that is available from neutron scattering experiments. The neutron structure factor, $S_N(q)$, has been calculated from the partial structure factors, $S_{\alpha\beta}(q)$, using the relation[3]

$$S_N(q) = \frac{N}{\sum_\alpha N_\alpha b_\alpha^2} \sum_{\alpha,\beta} b_\alpha b_\beta S_{\alpha\beta}(q) \qquad (7)$$

where $\alpha, \beta$ are the different atomic species, $b_\alpha$ is the coherent neutron-scattering length[71], $N_\alpha$ is the number of atoms for species $\alpha$ and $N$ is the total number of atoms. Figure 3 shows the neutron structure factor for alkali silicates compared to experimental data when available.

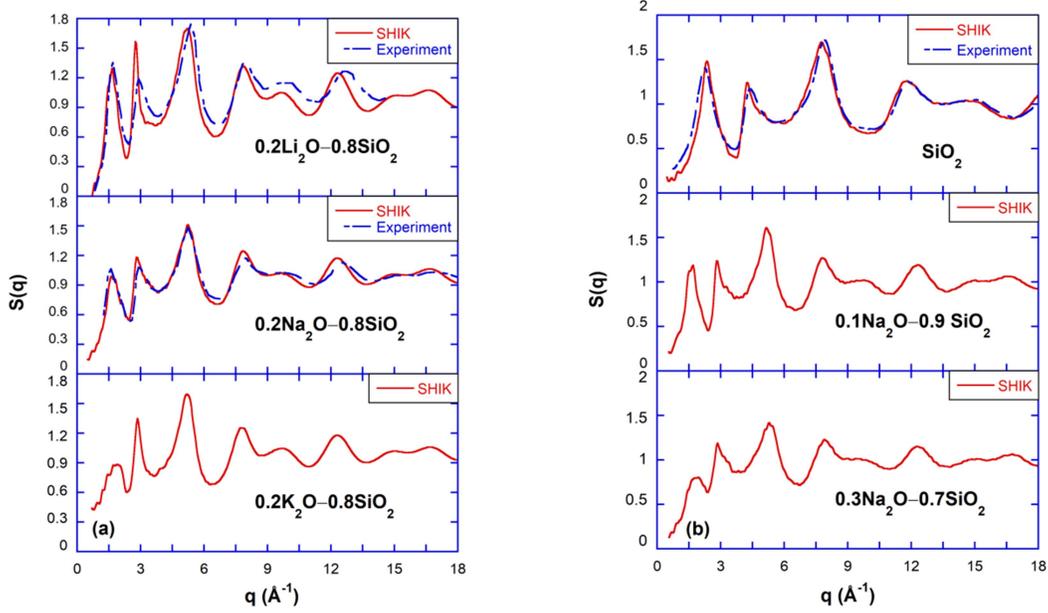

**Fig. 3** (a) Neutron structure factor as predicted by the SHIK potential (red solid line) and as measured in experiments[72–74] (blue dashed line) at ambient conditions. (a) Binary silicates of composition 0.2 $X_2O$–0.8$SiO_2$, where X is Li, Na, and K, (b) sodium silicate glasses with varying amount of modifier.

For the sodium and lithium silicate system, the first sharp diffraction peak reproduces very well the experimental data, indicating that the medium range order is described correctly by the SHIK potential[72–74]. Some discrepancies are noticeable for the lithium silicate at larger wave-vectors in that, e.g., the second peak predicted by the simulations is significantly higher than the experimental one. Such discrepancies are also observed at larger $q$, indicating that the structure on very small length scales is only qualitatively correct, but not quantitatively. Interestingly no such discrepancies are observed for the case of the sodium silicate in that for this system the predicted structure factor matches very well the experimental one.

These graphs also show that, as the alkali goes from Li to Na to K, the intensity of the first sharp peak, located at around 1.5 Å$^{-1}$, reduces and becomes broader. A similar effect is seen when the amount of modifier increases for a specific alkali silicate, see Fig. 3(b). Since a similar behavior was also observed for sodium silicate in experiments[72,73], we can conclude that our potential is indeed able to reproduce this effect.

Another useful way to analyze the medium range order of glass structure is to look at the primitive ring statistics[35,75]. The distribution of the size of the rings was calculated using the R.I.N.G.S code[76] and is shown in Fig. 4. Figure 4(a) shows that when 20% alkali is added to silica glass, the population of 6-8 membered rings decreases substantially, whereas the concentration of rings larger than 10 increases substantially, indicating that the network becomes increasingly disordered and depolymerized. As the alkali goes from Li to K, the number of 6-8 membered ring decreases, while the fraction of 5-membered rings increases and the same is found for larger rings. The same qualitative trend is seen if the concentration of modifier is increased for a specific alkali silicate glass, see Fig. 4(b), where the fraction of 5-membered rings and larger rings increase significantly at higher modifier contents.

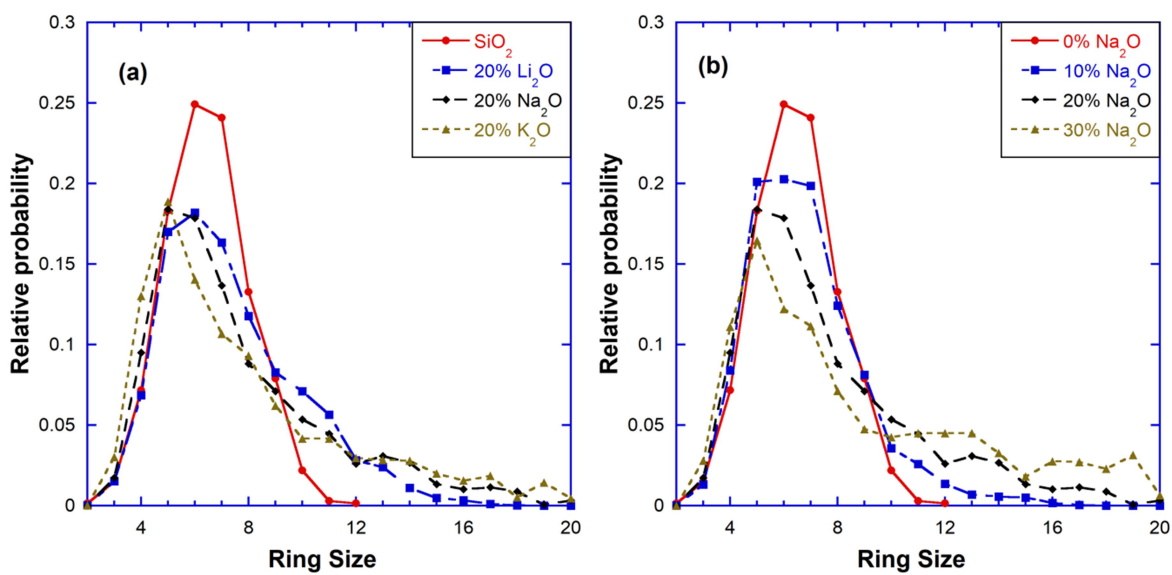

**Fig. 4** (a) Primitive ring statistics for (a) binary silicate glasses containing 20 mol% of alkali modifiers, and (b) sodium silicate glasses with different amount of modifier.

Since the addition of the alkali modifiers to silica results in the formation of non-bridging oxygen, it is of interest to quantify the resulting change in the structure in more detail. For this one can determine the $Q_n$ distributions where $n$ is the number of bridging oxygen per tetrahedral unit. Figure 5 shows the predicted $Q_n$ distribution for the various alkali silicate glasses as the amount of alkali is increased, in comparison to experimental data[77]. We see that the SHIK potential is able to reproduce very well the trends found in the experiments although for each alkali the amount of $Q_3$ is underestimated while the concentration of $Q_2$ and $Q_4$ are overestimated, especially at higher alkali content. Similar results have been observed when using other pair potentials[23,41] and usually are attributed to the much faster cooling rates used in MD simulations[41], which is likely to be also true in our case.

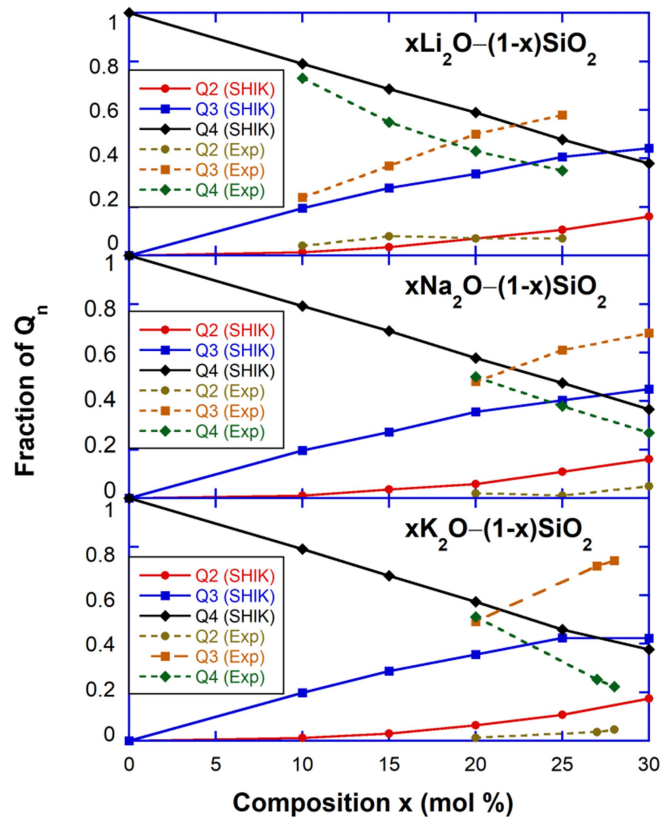

**Fig. 5** Fraction of $Q_n$ for various alkali silicates versus the amount of modifier as predicted by the SHIK potential, in comparison with results from experiments[77].

### 3.3 Properties of the glass

The simplest macroscopic quantity to compare with experiments is the glass density for various compositions. Figure 6(a) shows the composition-dependence of density at zero pressure and 300 K as the amount of alkali/alkaline-earth modifier is varied. The top two panels in Fig. 6(b) show the composition-dependence of the density in ternary systems with a constant amount of alkali as the amount of alumina is varied, while the bottom panel shows the density at a constant amount of silica as the amount of alkali is varied.

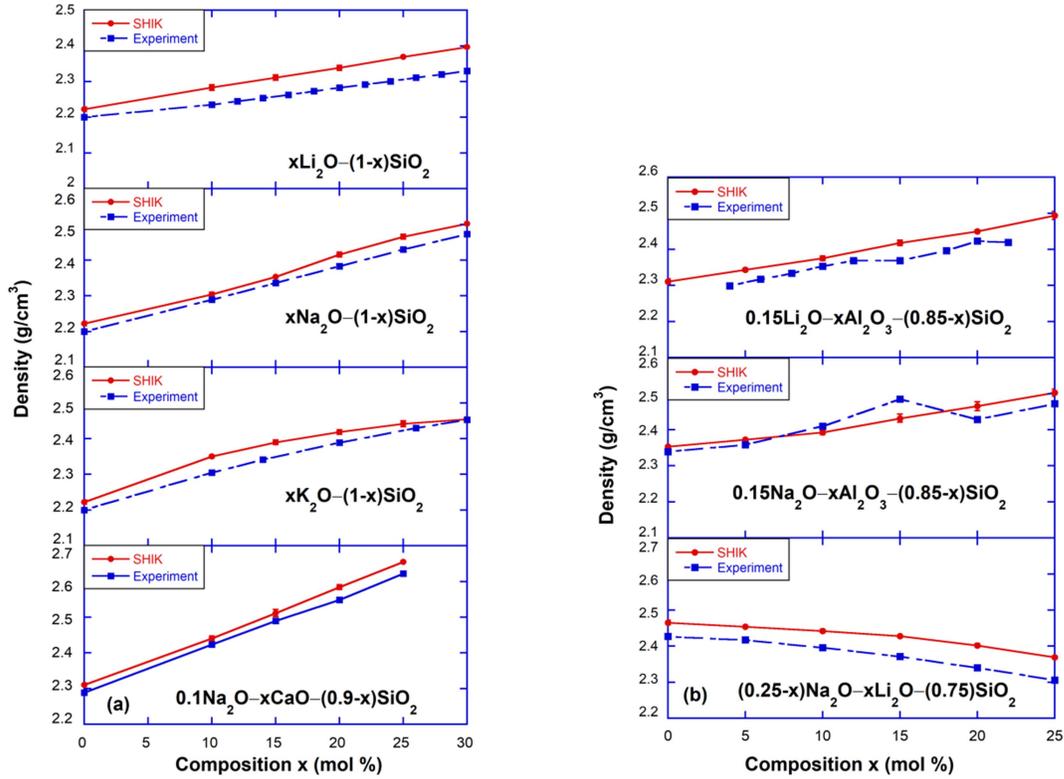

**Fig. 6** Density of (a) alkali silicates (top 3 panels) and sodium-calcium silicate (bottom panel) with varying modifier concentration, (b) sodium and lithium aluminosilicates with varying amount of Al at a constant modifier content (top 2 panels) and sodium-lithium silicate with varying ratios of modifiers at a constant amount of silica (bottom panel) for the SHIK potential (red solid line) compared to experimental data (blue dashed line)[4,78,79].

These graphs demonstrate that the new potentials are not only able to reproduce reliably the trends in density found in the experimental data when the concentration of the modifier is changed[4,78,79], they are also able to predict the density values within a few percent error over a wide range of compositions. Note that since for calcium silicate glass not much experimental data was available because of phase separation[80], we compare our data for the ternary system of sodium-calcium silicate instead, see bottom panel in Fig. 6(a). This result for the sodium-calcium silicate indicates that the SHIK potential is not only able to predict well properties of the binary silicates, but is also very reliable when different modifiers are mixed together. This is further exemplified in Fig. 6(b), where it can be seen that our potential predicts indeed very well the densities for various ternary silicates including the ones containing aluminum, see the top two panels. We mention that the short-range interactions with aluminum were optimized using a sodium aluminosilicate composition as the reference, but since aluminum does not require short-range interaction with the alkali, we can easily extend to compositions with other alkali from parameters that have already been optimized from the binary silicate systems. This can be seen in the top two panels of Fig. 6(b), which show that the new potentials are reliable even if the alkali type is changed.

A further quantity of great practical interest is the elastic moduli of glass since they are essential for a reliable description of the mechanical properties of the material. We note that most interaction potentials for silicate glasses are not able to predict correctly either the magnitude or the compositional dependence of the mechanical properties if the samples are not quenched in the NVT ensemble[23,41,46,47]. This flaw is avoided in the SHIK potential which is able to give a surprisingly good description of the various elastic moduli. Figure 7(a) and (b) show the bulk modulus and the Young's modulus, respectively, predicted for the various alkali and alkaline-earth compositions as the amount of modifier is increased.

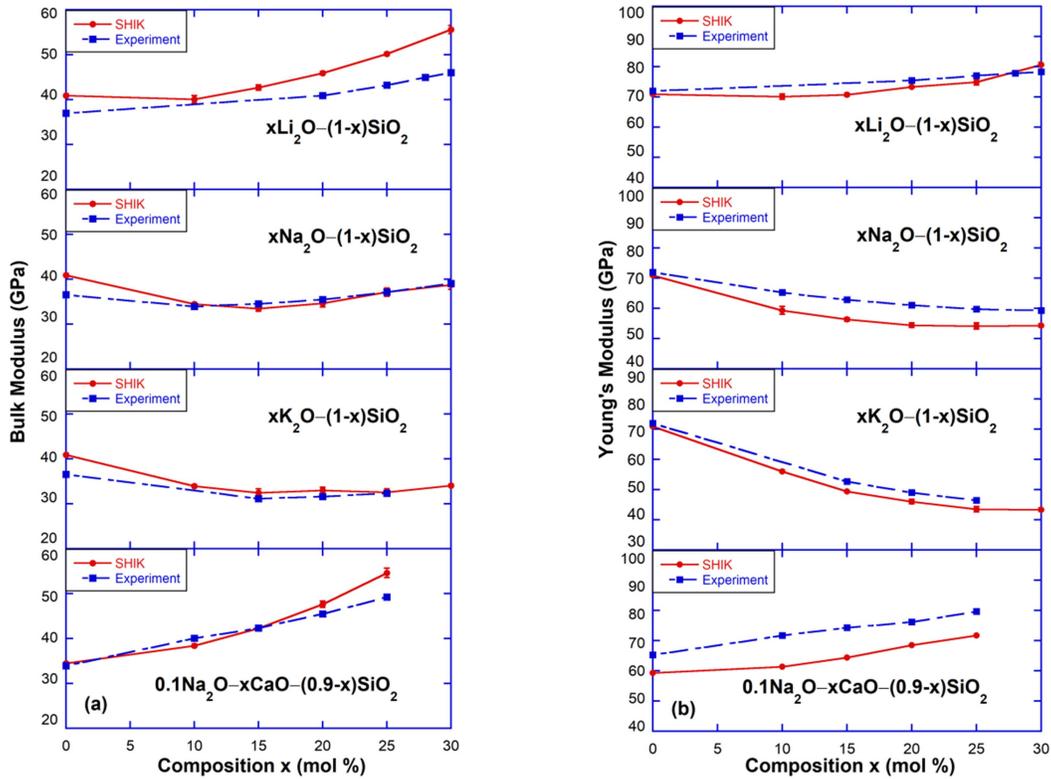

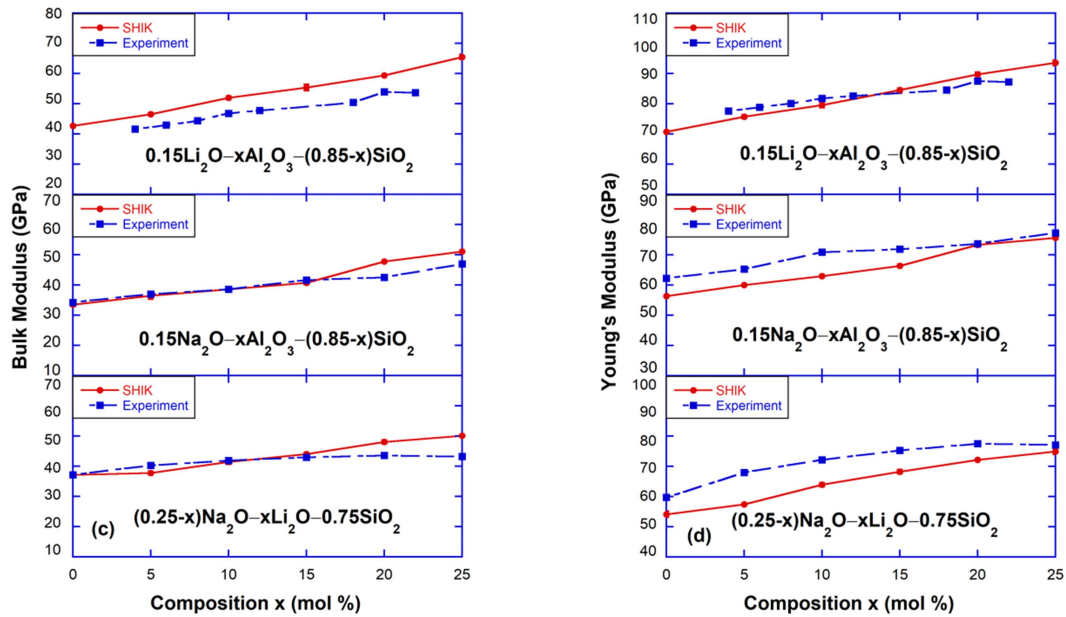

**Fig. 7** (a) Bulk modulus and (b) Young's modulus as a function of the modifier content for various alkali silicates and for sodium-calcium silicate at a constant sodium content. (c) Bulk modulus and (d) Young's modulus for alkali-aluminosilicate glasses (top 2 panels) at a constant modifier content and for sodium-lithium silicate at different modifier ratios and with a constant amount of silica (bottom panel). In all panels results from the SHIK potential are shown by the red solid lines and the experimental data are by the blue dashed lines[4,78,79].

Figure 7(a) shows that for sodium and potassium silicates the bulk modulus initially decreases and starts to increase again if more than 15 mol% of modifiers is added, while the Young's modulus in Fig. 7(b) decreases monotonically in the composition range studied in this work. This phenomenon has been attributed to the competition between the reduction in the elastic modulus due to the depolymerization of the silica network and the increase due to new bonds formed with the modifiers[81]. The lithium silicate system, top panels of Fig. 7(a) and (b), behaves differently due to the small size of lithium. Differences in the behavior of the bulk and Young's moduli with increasing alkali content for the different alkali systems can be attributed to the reduced free volume which results in an increased bulk modulus[41]. Figure 7(a) and (b) also demonstrate that the SHIK potential is not only able to predict the trends in the elastic moduli as a function of composition but also reproduce well their magnitude over the range of compositions studied here in that the predicted values differ at most by 10% from the experimental ones[4,78,79]. The top two panels in Fig. 7(c-d) show how these properties are predicted for lithium and sodium aluminosilicate compositions at a constant alkali content while the bottom panel shows how they vary when different alkali are mixed together in different ratios with a constant amount of silica. From these figures we recognize that the SHIK potential is not only accurate for the binary systems but also for the ternaries systems that include aluminum. This good agreement with experimental data demonstrates further the transferability and hence the predictive power of the new potential to explore new compositions.

Finally, we probe the ability of the potential to reproduce the vibrational density of states (VDOS) calculated from the Fourier transform of the velocity autocorrelation function[82]. We note that the most existing classical interaction potentials in the literature have not been able to reproduce the various features of the VDOS even for pure silica glass[13,16,41,83]. In our previous work, we hence included the VDOS in the cost function while optimizing the interaction parameters for silica. Even though the VDOS was not reproduced exactly, we observed a significant improvement over many other potentials, especially at lower and intermediate frequencies[6] of the VDOS. In Fig. 8 we compare the VDOS predicted by the SHIK potential with that from *ab initio* simulations for $0.4Na_2O–0.6SiO_2$[84]. One recognizes that the SHIK potential is indeed able to predict most of the features of the VDOS calculated from *ab initio* simulations, even if this type of observable is very different from the ones included in the cost-function of the optimization. Since there are some discrepancies observed in the intensities of the various peaks, we have tried if it is possible to remove them by including the VDOS in the cost function of the optimization. We found that this modification did not really improve the agreement, a result that is probably related to the fact that many of the vibrational features depend more strongly on the silica network than on the modifier. Hence in the end we did not include the VDOS in the cost function of the optimization in this work.

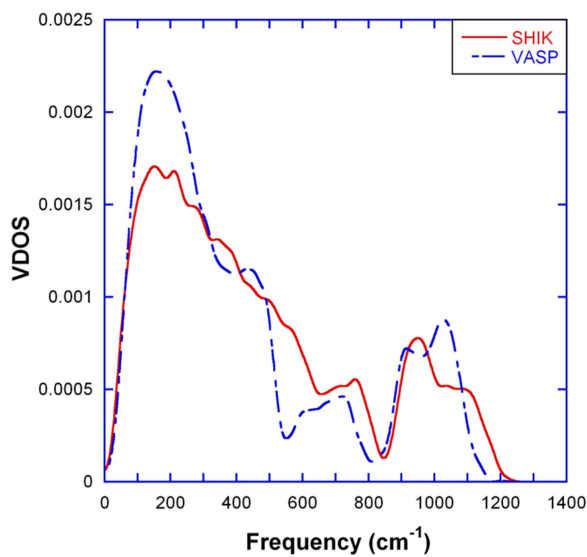

**Fig. 8** VDOS predicted by the SHIK potential (red solid line) compared to *ab initio* simulations (blue dashed line) for $40\%Na_2O–60\% SiO_2$.

## 4. Summary and Conclusions
In the present work we have used an optimization scheme similar to the one developed earlier to parameterize interaction potentials for alkali and alkaline-earth aluminosilicate glasses. For this we included in the cost function the melt structure at high temperature and the density and elastic modulus of the glass at room temperature. Parameters from the previous optimization for pure silica[5] were maintained to ensure transferability. The charge balancing scheme suggested by Habasaki et al.[25] to partially emulate the polarization effect was used for these potentials as it had shown the ability to predict well trends in density and mechanical properties for alkali silicates[46]. This simple scheme also allows high computational efficiency to study large and complex systems. This new set of interaction parameters is found to predict reliably both the trends and absolute values of different properties over a wide range of compositions for systems having a single type of alkali or a mixture of modifiers. This transferability enables therefore the easy exploration of structure and properties of new multi-component glasses not yet synthesized in experiments. Further improvements may be made to these interaction potentials by exploring both the functional form and the parameter space using advanced optimization techniques like machine learning[85–87].

## Acknowledgements
L. Huang acknowledges the financial support from the US National Science Foundation under grant No. DMR-1105238 and DMR-1255378, and a Corning-CFES seed fund, as well computational resources from the Center for Computational Innovations (CCI) at RPI. S. Ispas acknowledges HPC resources from GENCI (Grants A0010907572 and A0030907572). S. Sundararaman acknowledges an NSF-IMI travel grant from DMR-0844014 to France to initiate this collaborative work.